# Nothing is Certain but Doubt and Tests

John A McDermid OBE FREng
Department of Computer Science, University of York
Deramore Lane, York, YO10 5GH, UK
john.mcdermid@york.ac.uk

*Abstract*— Effective software safety standards will contribute to confidence, or assurance, in the safety of the systems in which the software is used. It is infeasible to demonstrate a correlation between standards and accidents, but there is an alternative view that makes standards "testable". Software projects are subject to uncertainty; good standards reduce uncertainty more than poor ones. Similarly assurance or integrity levels in standards should define an uncertainty gradient. The paper proposes an argument-based method of reasoning about uncertainty that can be used as a basis for conducting experiments (tests) to evaluate standards.

## I. INTRODUCTION

If software safety standards were effective then they should contribute to achievement or assurance of safety (or both). However, it is infeasible to show effectiveness of standards in terms of operational safety; the paper briefly examines the basis for this claim then outlines an approach to evaluation ("testing") of standards using arguments about uncertainty.

### A. The Unplanned Experiment

The motivation for the workshop explains the infeasibility of showing that software has, say, a $10^{-9}$ per hour unsafe failure rate [1]; thus standards are an "unplanned experiment". There are additional factors that emphasize the infeasibility of establishing a direct link between standards and safety:

- Cause-effect relationship – software development depends on the skills of engineers, management, tools, resources, etc.; showing the contribution of standards to achieved safety (cause and effect) is very difficult;
- Standards production – standards arise from a social process, do not reflect what any company actually does (some are closer than others), and are subject to (a high degree) of interpretation.

Arguably, these two factors (especially the latter) explain the significant differences observed in published standards [2].

### B. How Would we Know?

Several authors have proposed ways of improving the standards process, e.g. the "filter model" [3]. This paper takes the view that the fundamental challenge in assuring software contributions to system safety is management of uncertainty; in short assurance increases as uncertainty decreases.

More specifically, there are two sorts of uncertainty:

1. About faults or limitations in requirements;
2. About faults or limitations in implementation.

We refer to these as essential and accidental uncertainty, respectively. In a perfect world, a standard would remove all uncertainty giving absolute assurance in the requirements and implementation; in the real world, a better standard, or higher assurance/integrity level in a standard, will reduce uncertainty more than weaker standards/lower assurance/integrity levels.

## II. SOURCES OF UNCERTAINTY

To be able to model and reason about uncertainty requires an identification of sources of uncertainty; this is challenging, but can be approached based on the essential/accidental split.

### A. Essential Uncertainty

Essential uncertainty, *vis a vis* the software requirements, arises from limited understanding of the physics (or chemistry, etc.) of the embedding system, including sensors and actuators, and in the operational environment. For example, do we know:

"How the radar's detection capability can vary with the weather, particulates from volcanoes, etc."?

This is a source of uncertainty that can be analyzed relative to a system model, e.g. Parnas' 4-variable model [4].

### B. Accidental Uncertainty

Accidental uncertainty reflects limitations in development methods and in tools, etc. In the context of standards, the focus is on the requirements, recommendations, or objectives that the standard "promotes" and the residual uncertainties.

If we ask "uncertainty with respect to what" we can find a number of candidates, e.g. the "4+1 principles" [5]. The view adopted here, is that there are two core principles:

1. The safety requirements completely reflect the contribution of the software in the system to hazards.
2. The software meets the safety requirements.

Principle 1 covers both potential software "failure modes" and positive contributions, e.g. mitigations.

## III. REASONING ABOUT UNCERTAINTY

To analyze standards we propose to reason (argue) about their influence over uncertainty (this is different to Holloway's analysis [6] of DO178C [7] using GSN [8]); as we shall see below this gives a basis for some "planned experiments".

### A. Approach to Argumentation

Several notations have been proposed for safety cases, e.g. GSN that was based on work by Toulmin [9], see Fig. 1:

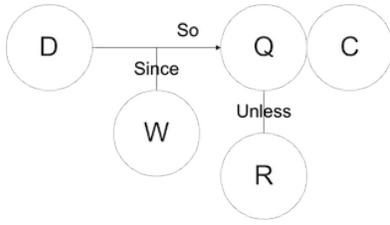

Fig. 1. Toulmin's Argument Structure

The elements of the structure are as follows:

- D – data (evidence) on which the argument rests;
- C – claim (conclusions) of the argument;
- W – warrant (justification) that the claim follows from the data;
- Q – qualification (uncertainty) of the claim;
- R – rebuttal, or source of the uncertainty.

Here we use Toulmin's structure as it addresses uncertainty and it is possible to "chain" arguments to counter the rebuttals – to argue that a particular source of uncertainty is controlled.

### B. Applying the Argumentation Approach

In applying the argumentation approach, we can consider how to reason about the sources of essential and accidental uncertainty. We briefly illustrate this with the principles set out at II B, relating mainly to accidental uncertainty. We note that many sources of uncertainty can affect a claim, so we employ a tabular form of argument, repeating argument elements, e.g. the qualification, as needed. The small fragment set out in Table 1 draws on the concepts in DO178C.

TABLE I. SAFETY REQUIREMENTS SATISFACTION

| Principle 2 | Software Safety Requirements Met | |
|---|---|---|
| | *Interpretation* | *Link* |
| C1 | All software safety requirements are satisifed by requirements testing and MC/DC testing. | |
| W | Testing shows requirements met under all path conditions | |
| Q1 | Subject to more complex data dependencies | R1 |
| Q2 | Subject to absence of run-time errors | R2 |

Arguments can be constructed to counter R1 and R2. For R2, we can carry out reviews against coding standards and for compatibility with the target machine (DO 178C objectives) or do static analysis with tools such as the SPARK Examiner [10].

Arguably use of tools such as the Examiner are better than human review, but both rest on further uncertainties – reviewer competence and tool integrity (assessed using the DO178C Tool Qualification Annex). You don't need the above approach in this simple case, but it (potentially) gives a systematic way of evaluating standards' "effectiveness" in all their complexity.

### IV. TOWARDS PLANNED EXPERIMENTS

The "plan for a plan" is simple; use the principles and argumentation approach to model standards of interest and then "test" the standards' effectiveness, for example by asking:

"Is the integrity/assurance level scheme in the standard valid in that higher levels reduce uncertainty more than lower ones?"

If it were valid, then the uncertainties would reduce with the level. This can be tested in several forms of experiment:

1. Paper/expert review;
2. Manufactured experiment, perhaps using students;
3. Real-world experience using company data.

These experiments become progressively more difficult, to conduct in a proper manner but doing the first would indicate whether or not doing the later ones would be worthwhile.

The author speculates that, if experimenting with standards, some, e.g. [11], would "fail" the above test, due to the way in which its requirements are stated; this might be a useful result in itself. Note that this is different to the approach in [12] that produces the "hidden" argument in standards; the approach set out here is using uncertainty as a "measure" of standards.

### V. CONCLUSIONS

This paper's hypothesis is that a constructive way to think about assessment of standards is via models of uncertainty, set out in argument form. It also suggests that the approach could potentially enable planned experiments to be carried out and it is believed that interesting insights would arise as a result.

Turning to the title: software development and assurance is beset by uncertainties (doubts). The discussion has focused on accidental uncertainties however testing always has a role in reducing the essential uncertainties (notwithstanding the use of simulation models), as only testing validates our understanding of the real world, including human operators.

REFERENCES

[1] R. W. Butler, G. B. Finelli, *The Infeasibility of Quantifying the Reliability of Life-Critical Real-*Time Software, IEEE Transactions on Software Engineering, 19(1), 1993.

[2] J. A. McDermid, D.J. Pumfrey, Software Safety: Why is there no concensus?, Proc ISSC, Hunstville AL, SSS, 2001.

[3] P. Steele, J. C. Knight, Analysis of Critical Systems Certification, Proc HASE, 2014, Miami Beach Fl, IEEE.

[4] D. L. Parnas, J. Madey, Functional Documents for Computer Programs, *Science of Computer Programming*, 25 (1), 1995.

[5] T.P. Kelly, Software Certification: Where is Confidence Won and Lost?, in Addressing Systems Safety Challenges, T. Anderson, C. Dale (Eds), Safety Critical Systems Club, 2014.

[6] DO178C, Software considerations in airborne systems and equipment certification.RTCA/DO-178C. RTCA, Inc., 2011.

[7] C. M. Holloway, Making the Implicit Explicit: Towards an Assurance Case for DO-178C, Proc ISSC, Boston, MA, ISSS, 2013.

[8] Goal Structuring Notation (GSN) community standard, see: http://www.goalstructuringnotation.info/documents/GSN_Standard.pdf.

[9] S. Toulmin, The Uses of Argument (updated), Cambrdige University Press, 2003.

[10] SPARK Examiner, see: http://www.adacore.com/sparkpro/ .

[11] CENELEC, EN 50128 – Railway applications – Communication, signalling and processing systems – Software for railway control and protection systems, 2011.

[12] P J. Graydon, T P. Kelly, Using argumentation to evaluate software assurance standards, Information & Software Technology 55(9): 1551-1562 (2013)